\documentclass[conference,a4paper]{IEEEtran}
\usepackage[a4paper, left=1.43cm, right=1.43cm, top=1.7cm, bottom=4.1cm]{geometry}
\usepackage[T1]{fontenc}
\usepackage[utf8]{inputenc}

\usepackage{amsmath}
\usepackage[cmintegrals]{newtxmath}
\usepackage{bm}
\usepackage{cite}
\usepackage{color}
\usepackage{graphicx}
\usepackage[acronym,toc,shortcuts]{glossaries-extra}
\setabbreviationstyle[acronym]{long-short}
\usepackage{siunitx}
\usepackage[hyphens]{url}
\usepackage{hyperref}
\usepackage[hyphenbreaks]{breakurl}
\usepackage{tablefootnote}
\usepackage{subcaption}
\usepackage{multirow}
\usepackage[official]{eurosym}
\usepackage{tikz}
\usetikzlibrary{positioning}
\usetikzlibrary{shapes}
\usetikzlibrary{spy}
\usetikzlibrary{calc}
\usetikzlibrary{arrows,automata}
\tikzset{>=latex}
\pgfdeclarelayer{bg}
\pgfsetlayers{bg,main}

\usepackage{pgfplots}
\pgfplotsset{compat=1.14}

\usepackage{array}
\newcolumntype{L}[1]{>{\raggedright\let\newline\\\arraybackslash\hspace{0pt}}m{#1}}
\newcolumntype{C}[1]{>{\centering\let\newline\\\arraybackslash\hspace{0pt}}m{#1}}
\newcolumntype{R}[1]{>{\raggedleft\let\newline\\\arraybackslash\hspace{0pt}}m{#1}}

\definecolor{middlegray}{rgb}{0.5,0.5,0.5}

\newacronym{ADC}{ADC}{analog to digital converter}
\newacronym{CMAC}{CMAC}{integrated 100\,Gbit/s Ethernet media access controller}
\newacronym{FPGA}{FPGA}{field programmable gate array}
\newacronym{HTG-ZRF8}{HTG-ZRF8-R2}{Hightech Global Xilinx Zynq\textregistered UltraScale+™ RFSoC Development Platform Rev. 2}
\newacronym{PCB}{PCB}{printed circuit board}
\newacronym{QSFP}{QSFP}{quad small formfactor pluggable transceiver}
\newacronym{SoC}{SoC}{system on chip}
\newacronym{VNA}{VNA}{vector network analyzer}
\newacronym{SDR}{SDR}{software-defined radio}
\newacronym{COTS}{COTS}{commercial off-the-shelf}
\newacronym{I/Q}{I/Q}{in-phase/quadrature}
\newacronym{RF}{RF}{radio frequency}
\newacronym{IRR}{IRR}{image rejection ratio}
\newacronym{USRP}{USRP}{universal software radio peripheral}
\newacronym{DSO}{DSO}{digital storage oscilloscope}
\newacronym{SINR}{SINR}{signal to interference and noise ratio}
\newacronym{PC}{PC}{personal computer}
\makeglossaries

\selectfont

\begin{document}


\makeatletter
\def\ps@IEEEtitlepagestyle{
  \def\@oddfoot{\mycopyrightnotice}
  \def\@evenfoot{}
}
\makeatother
\def\mycopyrightnotice{
  {\scriptsize
      \begin{minipage}{\textwidth}
      	\centering
      	Published in 2023 17th European Conference on Antennas and Propagation (EuCAP). DOI: 10.23919/EuCAP57121.2023.10133262\\
      	Coypright~\copyright~2023 IEEE.
      	Personal use of this material is permitted.
      	Permission from IEEE must be obtained for all other uses, in any current or future media, including reprinting/republishing this material for advertising or promotional purposes, creating new collective works, for resale or redistribution to servers or lists, or reuse of any copyrighted component of this work in other works.
      \end{minipage}
    }
}

\title{Receiver Bandwidth Extension Beyond Nyquist\\ Using Channel Bonding}

\author{\IEEEauthorblockN{Sebastian Giehl\IEEEauthorrefmark{1}, 
Carsten Andrich\IEEEauthorrefmark{1}\IEEEauthorrefmark{2}, Michael Schubert\IEEEauthorrefmark{2}, Maximilian Engelhardt\IEEEauthorrefmark{2}, Alexander Ihlow\IEEEauthorrefmark{1}}

\IEEEauthorblockA{\IEEEauthorrefmark{1}Technische Universität Ilmenau, Institute for Information Technology, Ilmenau, Germany}

\IEEEauthorblockA{\IEEEauthorrefmark{2}Fraunhofer Institute for Integrated Circuits IIS, Ilmenau, Germany}}

\maketitle

\begin{abstract}
  Current and upcoming communication and sensing technologies require ever larger bandwidths. 
  Channel bonding can be utilized to extend a receiver's instantaneous bandwidth beyond a single converter's Nyquist limit.  
  Two potential joint front-end and converter design approaches are theoretically introduced, realized and evaluated in this paper. 
  The Xilinx RFSoC platform with its 5\,GSa/s \acp{ADC} is used to implement both a hybrid coupler based \ac{I/Q} sampling and a time-interleaved sampling approach along with channel bonding.
  Both realizations are demonstrated to be able to reconstruct instantaneous bandwidths of 5\,GHz with up to 49\,dB \ac{IRR} typically within 4 to 8\,dB the front-ends' theoretical limits.
\end{abstract}

\begin{IEEEkeywords}
  Bandwidth extension, channel bonding, I/Q sampling, hybrid coupler, sample interleaving.
\end{IEEEkeywords}

\IEEEpeerreviewmaketitle
\glsresetall
\section{Introduction}
Modern communication and sensing technologies include mobile communications, array signal processing, beamforming, localization as well as joint communication and sensing approaches. 
To cover the widest possible range of applications, transceiver hardware for research and development in this field is required to be capable of both an extendable number of channels and an ultra-high instantaneous bandwidth or time resolution.
While \acp{DSO} may offer superior sampling rates, they are not capable of performing continuous sampling~\cite{keysightUXR}.
At the same time, standalone \ac{SDR} devices such as \acp{USRP} are constrained by their holistic system design~\cite{x410spec}.

\begin{figure}[h]
  \centering
  \resizebox{.95\linewidth}{!}{
    \begin{tikzpicture}[->,>=stealth', transform shape]
      \node at(5,0) [draw,rectangle, minimum height=3cm, minimum width= 2cm, align = center] (b) {Analog\\Front-End};
      \node at(9,0) [draw,rectangle, minimum height=4cm, minimum width= 5cm,label={[label distance=-.5cm]270:HTG-ZRF8 Board}] (c) {};
      \node at(7.75,.75) [draw,rectangle,minimum height=1.5cm, minimum width= 1.5cm, align= center] (d) {8x\\DAC\\10\,GSa/s};
      \node at(7.75,-.75) [draw,rectangle,minimum height=1.5cm, minimum width= 1.5cm, align= center] (e) {8x\\ADC\\5\,GSa/s};
      \node at(10,0) [draw,rectangle,minimum height=3cm, minimum width= 2cm, text width = 1.5cm, align= center] (f) {FPGA ZU48DR};
      \node at(15.25,0) [draw,rectangle, minimum height=3cm, minimum width= 2.5cm, align = center] (h) {Host};
      \node at(15.25,1) [draw,rectangle, minimum height=1cm, minimum width= 2.2cm] (i) {Processing};
      \node at(15.25,-1) [draw,rectangle, minimum height=1cm, minimum width= 2.2cm] (j) {Recording};

      \path
      (d) 		edge[line width=1.2pt]   (b.east|-d)
      (b.east|-e) 		edge[line width=1.2pt]   (e)
      (e) 		edge[line width=1.2pt]   (f.west|-e)
      (f.west|-d) 		edge[line width=1.2pt]   (d)
      (f) 		edge[<->, double,line width=1.2pt] node[above, align=center]{100\,Gbit/s\\Ethernet}  (h)
      ;

    \end{tikzpicture}
  }
  \caption{\glsfmtshort{SoC} carrier board HTG-ZRF8 in full duplex measurement setup connected to a host server using 100\,Gbit/s Ethernet for sample recording and processing~\cite{HTG-Site}.}
  \label{fig:targetsystemblocks}
\end{figure}
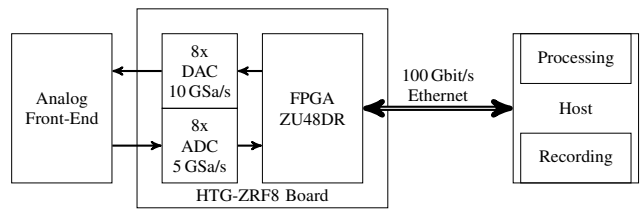

As a solution, \ac{FPGA}-based \acp{SoC} with integrated direct \ac{RF} sampling capable data converters like Xilinx' RFSoC series allow realizing versatile measurement platforms cost-efficiently~\cite{ds889}.
Figure~\ref{fig:targetsystemblocks} illustrates an exemplary measurement system's setup. 
Beside the data converter and its analog front-end, a powerful, high-speed data interface is required to connect the converter to a host server for sample recording or processing.
With \ac{ADC} sample rates of up to 5\,GSa/s, the RFSoC series is able to handle signals with continuous bandwidths of up to 2.5\,GHz on each of its eight channels. 
At the same time, it integrates two hardware blocks each realizing a 100\,Gbit/s Ethernet high-speed data interface which can be used to implement sample streaming.
Although the RFSoC already offers high sampling rates, further extension of the instantaneous bandwidth beyond the Nyquist limit of an individual channel is desired~\cite{Landau}.

To overcome the Nyquist limit for individual converters, channel bonding is utilized to reconstruct an enhanced instantaneous bandwidth from several converter channel's data.
This paper introduces two front-end architectures for \ac{ADC} channel bonding and evaluates their performance using an \ac{FPGA} design which realizes sample streaming over 100\,Gbit/s Ethernet.

\section{State of the Art}
Common \ac{SDR} solutions like \acp{USRP} implement homodyne or heterodyne \ac{I/Q} mixing front-ends to cover an extended bandwidth.
Beside incurring an additional synchronization effort, the front-end limits the instantaneous bandwidth to several hundred MHz.~\cite{x410spec}

Exploiting the bandpass Nyquist theorem, implicit downmixing can be performed by real-valued subsampling as shown in~\cite{undersampling1}.
In addition, \cite{undersampling2} uses spectral sparsity properties to further lower the required sampling rate or the requirements for the input band filter. 
Nevertheless, due to real-valued sampling -- in contrast to the \ac{I/Q} mixing -- the signal band must not cross an integer multiple of the Nyquist frequency and is therefore constrained by the converter's sampling rate.

In order to overcome the Nyquist limit of an individual data converter, channel bonding has to be introduced.
Interleaved sampling is a widespread bonding principle. 
It implements time shifted sampling and sample multiplexing on multiple channels and can be found in \ac{COTS} high rate \acp{ADC} or built based on separate converters as demonstrated by \cite{interleaved1} or \cite{snaa286} and shown below. 

\section{Signal Theory}
Sampling in time leads to periodification in the frequency domain and therefore might introduce aliasing per channel, which is described by the Nyquist criterion~\cite{Landau}.

Channel bonding is intended to extend the instantly receivable signal bandwidth beyond the Nyquist bandwidth of a single converter channel by utilizing specific front-end architectures and multiple \ac{ADC} channels. 
An ideal approach shall provide a reconstructed spectrum with flat frequency response and without any inband aliasing.
Each bonding technique requires a specific inter-channel relation.
While in theory, reconstruction realizes perfect alias cancellation, real implementations' non-idealities may limit the achievable image attenuation.
The achieved \ac{IRR} is the central quality factor of a front-end architecture. 

\begin{figure}
  \centering
  \begin{subfigure}[b]{\linewidth}
    \centering
    \resizebox{.99\linewidth}{!}{
      \begin{tikzpicture}\large

        \draw[->] (-6.5, 0) -- (6.5, 0) node[right] {$f\text{(GHz)}$};
        \draw[->] (0, -.1) -- (0,1.6) node[above] {$\left| {H_\text{Filterbank, attenuated}(f)}\right|$};
        \draw (-6,.1)-- ++(0,-0.2) node[below]{-6} ++(1,0.2)-- ++(0,-0.2) node[below]{-5} ++(1,0.2)-- ++(0,-0.2) node[below]{-4} ++(1,0.2)-- ++(0,-0.2) node[below]{-3} ++(1,0.2)-- ++(0,-0.2) node[below]{-2} ++(1,0.2)-- ++(0,-0.2) node[below]{-1} ++(1,0.2)-- ++(0,-0.2) node[below]{0} ++(1,0.2)-- ++(0,-0.2) node[below]{1} ++(1,0.2)-- ++(0,-0.2) node[below]{2} ++(1,0.2)-- ++(0,-0.2) node[below]{3} ++(1,0.2)-- ++(0,-0.2) node[below]{4} ++(1,0.2)-- ++(0,-0.2) node[below]{5} ++(1,0.2)-- ++(0,-0.2) node[below]{6};
        \draw (-.1,1)node[above left]{1}--++(.2,0) ++(-.2,-.5)--++(.2,0);

        \draw[line width = 2pt,green](2.5,0)--++(-1,1)--++(-3,0)--++(-1,-1);

        \draw[line width = 2pt,blue](5,0)--++(-1,1)--++(-.5,0)--++(-1,-1);
        \draw[line width = 2pt,blue](-2.5,0)--++(-1,1)--++(-.5,0)--++(-1,-1);

        \draw[line width = 2pt, dashed](2.5,1.4)node[right]{$+f_\text{Ny}$} --++(0,-1.6) ;
        \draw[line width = 2pt, dashed](-2.5,1.4)node[left]{$-f_\text{Ny}$} --++(0,-1.6) ;
        \draw[line width = 2pt, dashed](5,1.4)node[right]{$+2 f_\text{Ny}$} --++(0,-1.6) ;
        \draw[line width = 2pt, dashed](-5,1.4)node[left]{$-2 f_\text{Ny}$} --++(0,-1.6) ;

      \end{tikzpicture}
    }
    \caption{Filterbank configuration without aliasing.}
    \label{fig:FilterbankAttenuation}
  \end{subfigure}
  \begin{subfigure}[b]{\linewidth}\centering
    \resizebox{.99\linewidth}{!}{
      \begin{tikzpicture}\large

        \draw[->] (-6.5, 0) -- (6.5, 0) node[right] {$f\text{(GHz)}$};
        \draw[->] (0, -.1) -- (0,1.6) node[above] {$\left| {H_\text{Filterbank, aliased}(f)}\right|$};
        \draw (-6,.1)-- ++(0,-0.2) node[below]{-6} ++(1,0.2)-- ++(0,-0.2) node[below]{-5} ++(1,0.2)-- ++(0,-0.2) node[below]{-4} ++(1,0.2)-- ++(0,-0.2) node[below]{-3} ++(1,0.2)-- ++(0,-0.2) node[below]{-2} ++(1,0.2)-- ++(0,-0.2) node[below]{-1} ++(1,0.2)-- ++(0,-0.2) node[below]{0} ++(1,0.2)-- ++(0,-0.2) node[below]{1} ++(1,0.2)-- ++(0,-0.2) node[below]{2} ++(1,0.2)-- ++(0,-0.2) node[below]{3} ++(1,0.2)-- ++(0,-0.2) node[below]{4} ++(1,0.2)-- ++(0,-0.2) node[below]{5} ++(1,0.2)-- ++(0,-0.2) node[below]{6};
        \draw (-.1,1)node[above left]{1}--++(.2,0) ++(-.2,-.5)--++(.2,0);

        \draw[line width = 2pt,green](3.5,0)--++(-1,1)--++(-5,0)--++(-1,-1);

        \draw[line width = 2pt,blue](6,0)--++(-1,1)--++(-2.5,0)--++(-1,-1);
        \draw[line width = 2pt,blue](-1.5,0)--++(-1,1)--++(-2.5,0)--++(-1,-1);

        \draw[line width = 2pt, dashed](2.5,1.4)node[right]{$+f_\text{Ny}$} --++(0,-1.6) ;
        \draw[line width = 2pt, dashed](-2.5,1.4)node[left]{$-f_\text{Ny}$} --++(0,-1.6) ;
        \draw[line width = 2pt, dashed](5,1.4)node[right]{$+2 f_\text{Ny}$} --++(0,-1.6) ;
        \draw[line width = 2pt, dashed](-5,1.4)node[left]{$-2 f_\text{Ny}$} --++(0,-1.6) ;

      \end{tikzpicture}
    }
    \caption{Filterbank configuration with continuous pass-band amplitude.}
    \label{fig:FilterbankAliasing}
  \end{subfigure}
  \caption{Filter bank comparison for real-valued sampling. Green: Lowpass. Blue: Bandpass. Neither approach enables flat pass-band magnitude while suppressing aliasing.}
  \label{fig:tradSamplingBandfilters}
\end{figure}
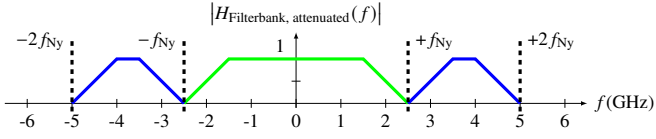
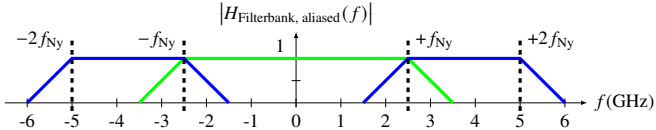

Figure~\ref{fig:tradSamplingBandfilters} visualizes two simple channel bonding setups using filter banks and real-valued sampling. 
While no flat amplitude response can be achieved implementing the filter bank shown in Fig.~\ref{fig:FilterbankAttenuation}, aliasing will be present in the signal sampled subsequently to the filter bank of Fig.~\ref{fig:FilterbankAliasing}.
Using individual channel real-valued sampling, a signal to be converted must not cross the Nyquist frequency to prevent aliasing, but at the same time, the signal must pass the filter unattenuated to preserve the desired flat amplitude response.
Since no infinitely steep filter edges can be realized, approaches like this can never meet both design goals specified above.
Instead, advanced receiver front-end approaches have to be utilized.

The Nyquist theorem for real-valued, sampled signals restricts the filter band's positions and therefore limits the ability to reconstruct a continuous signal spectrum from band-filtered signals.
Introducing \ac{I/Q} sampling leads to an analytic signal representation and therefore allows arbitrary positioning of the individual filter bands without incurring aliasing.
As an alternative, using time interleaved sampling continuous spectral reconstruction can be performed.

\subsection{I/Q sampling hybrid coupler approach}
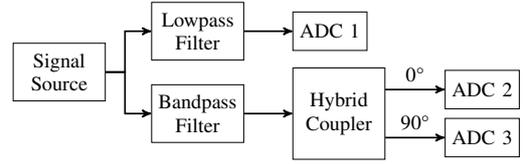
\begin{figure}[h]
  \centering
  \resizebox{.77\linewidth}{!}{
    \begin{tikzpicture}[->,>=stealth', transform shape]\large
      \node at (0,0)[rectangle,draw= black, minimum height = 1.2cm, minimum width = 1.9cm, align = center](Source){Signal\\Source};
      \node at($(Source.east)+(1.9,.85)$)[rectangle, draw =black,minimum height =1.2cm, minimum width=1.9cm , align = center,inner sep = 0mm](LPF){Lowpass\\Filter};
      \node [below = .5cm of LPF][rectangle, draw =black,minimum height = 1.2cm, minimum width=1.9cm , align = center,inner sep = 0mm](BPF){Bandpass\\Filter};

      \node [right = 1cm of BPF] [rectangle, draw =black,minimum height = 1.9cm, minimum width=1.9cm , align = center](hyb){Hybrid\\Coupler};

      \node [right= 1cm of LPF] [rectangle, draw =black,minimum height = .8cm, minimum width=1.5cm , align = center](ADC1){ADC 1};

      \node at($(hyb.east)+(2,.5)$) [rectangle, draw =black,minimum height = .8cm, minimum width=1.5cm , align = center](ADC2){ADC 2};

      \node at($(hyb.east)+(2,-.5)$) [rectangle, draw =black,minimum height = .8cm, minimum width=1.5cm , align = center](ADC3){ADC 3};

      \draw[->,line width = 1pt](Source.east) --++ (.4,0) |- (LPF);
      \draw[->,line width = 1pt](Source.east) --++ (.4,0) |- (BPF);
      \path(LPF) edge [->,line width = 1pt] (ADC1)
      (BPF) edge [->,line width = 1pt] (hyb)
      (hyb.east |- ADC2) edge [->,line width = 1pt]node[midway, above]{0°} (ADC2)
      (hyb.east |- ADC3) edge [->,line width = 1pt]node[midway, above]{90°} (ADC3)
      ;

    \end{tikzpicture}
  }
  \caption{\glsfmtshort{I/Q} sampling: Functional blocks.}
  \label{fig:hyb90sys}
\end{figure}

Using an analytic signal representation overcomes the limitations of the filter bank approach with real-valued sampling.
The baseband signal path shown in Fig.~\ref{fig:hyb90sys} can still be sampled real-valued by a single converter.
In contrast to the approaches shown in Fig.~\ref{fig:tradSamplingBandfilters}, performing \ac{I/Q} sampling of the bandpass signal by utilizing a 90° hybrid coupler and two coherent converter channels allows the filter to be designed seamlessly aligned. 

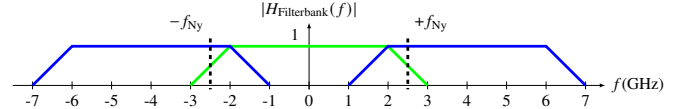
\begin{figure}
  \centering
  \resizebox{.99\linewidth}{!}{
    \begin{tikzpicture}\large
      \coordinate (shift_Xin) at (0,0);
      \coordinate (shift_Hsamp) at (0,-5);
      \coordinate (shift_Hyb) at (0,-10);

      \begin{scope}[shift=(shift_Hyb)]
        \draw[->] (-7.5, 0) -- (7.5, 0) node[right] {$f\text{(GHz)}$};
        \draw[->] (0, -.1) -- (0,1.6) node[above] {$\left| {H_\text{Filterbank}(f)}\right|$};
        \draw (-7,.1)-- ++(0,-0.2) node[below]{-7} ++(1,0.2)-- ++(0,-0.2) node[below]{-6} ++(1,0.2)-- ++(0,-0.2) node[below]{-5} ++(1,0.2)-- ++(0,-0.2) node[below]{-4} ++(1,0.2)-- ++(0,-0.2) node[below]{-3} ++(1,0.2)-- ++(0,-0.2) node[below]{-2} ++(1,0.2)-- ++(0,-0.2) node[below]{-1} ++(1,0.2)-- ++(0,-0.2) node[below]{0} ++(1,0.2)-- ++(0,-0.2) node[below]{1} ++(1,0.2)-- ++(0,-0.2) node[below]{2} ++(1,0.2)-- ++(0,-0.2) node[below]{3} ++(1,0.2)-- ++(0,-0.2) node[below]{4} ++(1,0.2)-- ++(0,-0.2) node[below]{5} ++(1,0.2)-- ++(0,-0.2) node[below]{6} ++(1,0.2)-- ++(0,-0.2) node[below]{7};
        \draw (-.1,1)node[above left]{1}--++(.2,0);

        \draw[line width = 2pt,green](3,0)--++(-1,1)--++(-4,0)--++(-1,-1);

        \draw[line width = 2pt,blue](7,0)--++(-1,1)--++(-4,0)--++(-1,-1);
        \draw[line width = 2pt,blue](-1,0)--++(-1,1)--++(-4,0)--++(-1,-1);

        \draw[line width = 2pt, dashed](2.5,1.2)node[above right]{$+f_\text{Ny}$} --++(0,-1.3) ;
        \draw[line width = 2pt, dashed](-2.5,1.2)node[above left]{$-f_\text{Ny}$} --++(0,-1.3) ;
      \end{scope}

    \end{tikzpicture}
  }
  \caption{\glsfmtshort{I/Q} sampling: Sampling filter bank. Green: Baseband, sampled real-valued. Blue: Bandpass, \glsfmtshort{I/Q} sampled. Enables flat pass-band magnitude and alias suppression.}
  \label{fig:hybFilterConfiguration}
\end{figure}

Due to the analytic representation of the bandpass signal, the Nyquist criterion then only limits the overall bandwidth, but not the absolute band position.
This allows to gain an overall flat amplitude response after reconstruction as Fig.~\ref{fig:hybFilterConfiguration} illustrates. 

An \ac{I/Q} signal consists of a real and an imaginary-valued part.
When generating the \ac{I/Q} signal, using a mixing analog front-end might seem convenient, but will increase the system's complexity.
Instead, an ultra wideband 90° hybrid coupler shall be used to generate both the in-phase and quadrature signals.
As a passive component to be utilized in-line, it does not require any additional components or circuitry.

Throughout its operating range, it is able to generate a 90° phase shifted version together with the 0° version at its outputs.
Coherently sampling both signals allows the reconstruction of a single sideband spectrum in the digital domain as shown:
\begin{equation}\label{eq:hybridF}
  \begin{split}
    y_{\text{BPF, analytic}}(t) &= y_{\text{BPF}}(t) + \text{j\,} (y_{\text{BPF, $\angle$}}(t))\\
    &= y_{\text{BPF}}(t) + \text{j\,} \mathscr{H}(y_{\text{BPF}}(t))\\\\
    Y_{\text{BPF, analytic}}(f) &= Y_{\text{BPF}}(f) + \text{j\,} Y_{\text{BPF, $\angle$}}(f)\\
    &= Y_{\text{BPF}}(f) + \text{j\,}(-\text{j} \text{\,sgn}(f) Y_{\text{BPF}}(f))\\
    &= (1+\text{sgn}(f)) Y_{\text{BPF}}(f)
  \end{split}
\end{equation}

In theory, the 90° phase shift can be expressed as the Hilbert transformation of the bandpass signal $y_{\text{BPF}}(t)$.
Interpreting the 0° signal $y_{\text{BPF}}(t)$ as real and the 90° signal $y_{\text{BPF, $\angle$}}(t)$ as imaginary-valued, the sum of both is the desired single sideband signal.
Designing the filter's pass band and transition band according to the channels' sampling rate allows to fully prevent aliasing within the reconstructed band.

\begin{figure}
  \centering
  \resizebox{\linewidth}{!}{
    \begin{tikzpicture}\large
      \coordinate (shift_LPF) at (-4.6,-5);
      \coordinate (shift_BPF) at (4.6,-5);
      \coordinate (shift_Hhyb) at (0,-11);
      \begin{scope}[shift=(shift_LPF)]
        \draw[->] (-3.5, 0) -- (3.8, 0) node[right] {$f\text{(GHz)}$};
        \draw[->] (0, -.1) -- (0,1.5) node[above] {$\text{Re}(Y_\text{BPF}(f))$};
        \draw (-3,.1) -- ++(0,-0.2) node[below]{-6} ++(.5,0.2)-- ++(0,-0.2) node[below]{-5} ++(.5,0.2)-- ++(0,-0.2) node[below]{-4} ++(.5,0.2)-- ++(0,-0.2) node[below]{-3} ++(.5,0.2)-- ++(0,-0.2) node[below]{-2} ++(.5,0.2)-- ++(0,-0.2) node[below]{-1} ++(.5,0.2)-- ++(0,-0.2) node[below]{0} ++(.5,0.2)-- ++(0,-0.2) node[below]{1} ++(.5,0.2)-- ++(0,-0.2) node[below]{2} ++(.5,0.2)-- ++(0,-0.2) node[below]{3} ++(.5,0.2)-- ++(0,-0.2) node[below]{4} ++(.5,0.2)-- ++(0,-0.2) node[below]{5} ++(.5,0.2)-- ++(0,-0.2) node[below]{6} ++(.5,0.2)-- ++(0,-0.2) node[below]{7};
        \draw (-.1,1)node[left]{1}--++(.2,0);
        \draw[line width = 2pt,blue](-3.5,0)--++(.5,1)--(-1,1)--++(.5,-1);
        \draw[line width = 2pt,blue](3.5,0)--++(-.5,1)--(1,1)--++(-.5,-1);
      \end{scope}
      \begin{scope}[shift=(shift_BPF)]
        \draw[->] (-3.5, 0) -- (3.8, 0) node[right] {$f\text{(GHz)}$};
        \draw[->] (0, -.1) -- (0,1.5) node[above] {$\text{Im}(Y_\text{BPF,~$\angle$}(f))$};
        \draw (-3,.1) -- ++(0,-0.2) node[below]{-6} ++(.5,0.2)-- ++(0,-0.2) node[below]{-5} ++(.5,0.2)-- ++(0,-0.2) node[below]{-4} ++(.5,0.2)-- ++(0,-0.2) node[below]{-3} ++(.5,0.2)-- ++(0,-0.2) node[below]{-2} ++(.5,0.2)-- ++(0,-0.2) node[below]{-1} ++(.5,0.2)-- ++(0,-0.2) node[below]{0} ++(.5,0.2)-- ++(0,-0.2) node[below]{1} ++(.5,0.2)-- ++(0,-0.2) node[below]{2} ++(.5,0.2)-- ++(0,-0.2) node[below]{3} ++(.5,0.2)-- ++(0,-0.2) node[below]{4} ++(.5,0.2)-- ++(0,-0.2) node[below]{5} ++(.5,0.2)-- ++(0,-0.2) node[below]{6} ++(.5,0.2)-- ++(0,-0.2) node[below]{7};
        \draw (-.1,1)node[left]{1}--++(.2,0);
        \draw[line width = 2pt,blue](-3.5,0)--++(.5,1)--(-1,1)--++(.5,-1);
        \draw[line width = 2pt,blue](3.5,0)--++(-.5,-1)--(1,-1)--++(-.5,1);
      \end{scope}

      \begin{scope}[shift=(shift_Hhyb)]
        \draw[->] (-3.5, 0) -- (3.8, 0) node[right] {$f\text{(GHz)}$};
        \draw[->] (0, -.1) -- (0,2.5) node[above] {$\left| {Y_\text{BPF,~analytic}(f)}\right|$};
        \draw (-3,.1) -- ++(0,-0.2) node[below]{-6} ++(.5,0.2)-- ++(0,-0.2) node[below]{-5} ++(.5,0.2)-- ++(0,-0.2) node[below]{-4} ++(.5,0.2)-- ++(0,-0.2) node[below]{-3} ++(.5,0.2)-- ++(0,-0.2) node[below]{-2} ++(.5,0.2)-- ++(0,-0.2) node[below]{-1} ++(.5,0.2)-- ++(0,-0.2) node[below]{0} ++(.5,0.2)-- ++(0,-0.2) node[below]{1} ++(.5,0.2)-- ++(0,-0.2) node[below]{2} ++(.5,0.2)-- ++(0,-0.2) node[below]{3} ++(.5,0.2)-- ++(0,-0.2) node[below]{4} ++(.5,0.2)-- ++(0,-0.2) node[below]{5} ++(.5,0.2)-- ++(0,-0.2) node[below]{6} ++(.5,0.2)-- ++(0,-0.2) node[below]{7};
        \draw (-.1,2)node[above left]{2}--++(.2,0) ++(-.2,-1)node[above left]{1}--++(.2,0);
        \draw[line width = 2pt,blue](-3.5,0)--++(.5,.2)--(-1,.2)--++(.5,-.2);
        \draw[line width = 2pt,blue](3.5,0)--++(-.5,2)--(1,2)--++(-.5,-2);

        \draw[line width = 2pt,dashed] (1,2) --++(-4.75,0);
        \draw[line width = 2pt,dashed] (-1,.2) --++(-2.75,0);
        \draw[<->] (-3.5,2) --++ (0,-0.9) node[right]{IRR}--++ (0,-0.9);
      \end{scope}

      \node at (-4.6,-7) [draw,minimum height = .8cm, minimum width = 1.5cm, rotate = 90,inner sep=0pt, outer sep=0pt, align = center](ADC2){ADC 2};
      \draw[<-, line width = 2pt](ADC2.east)--++(0,.5);

      \node at (4.6,-7) [draw,minimum height = .8cm, minimum width = 1.5cm, rotate = 90,inner sep=0pt, outer sep=0pt, align = center](ADC3){ADC 3};
      \draw[<-, line width = 2pt](ADC3.east)--++(0,.5);

      \node at (0,-7) [circle, draw,minimum  size = 15pt,inner sep=0pt, outer sep=0pt, align = center](mult3){};

      \draw[<-, line width = 2pt](mult3)--++(1,0)--node[above]{j}++(1,0);
      \draw[<-, line width = 2pt](mult3)--++(-2,0);
      \draw[->, line width = 2pt](mult3)--++(0,-.8);
      \draw(mult3)++(0.2,0)--++(-0.4,0)++(0.2,0.2)--++(0,-0.4);
    \end{tikzpicture}
  }
  \caption{\glsfmtshort{I/Q} sampling: Analytic signal generation from 0° and 90° signals coherently sampled on two converter channels.}
  \label{fig:hyb90rxanalytic}
  \vspace{-.3cm}
\end{figure}
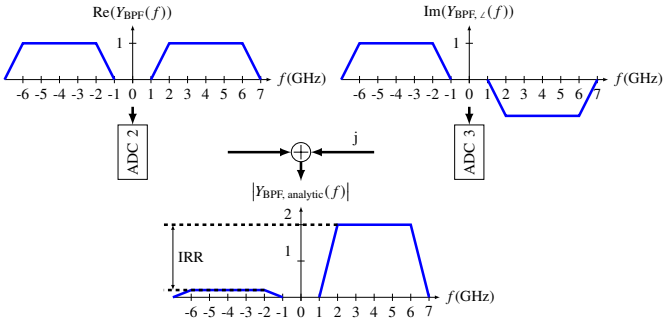

Figure~\ref{fig:hyb90rxanalytic} visualizes the effect of the reconstruction to the signal's sidebands.
Hereby, a real implementation's non-idealities as phase or amplitude imbalances may introduce a finite \ac{IRR}.

\subsection{Interleaved sampling}
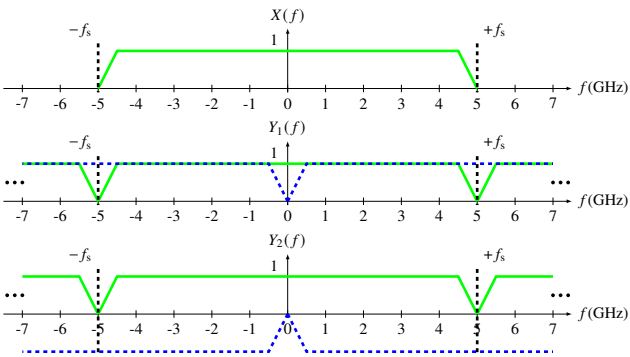
\begin{figure}[h]\centering
  \vspace{-.2cm}
  \resizebox{.95\linewidth}{!}{
    \begin{tikzpicture}\large
      
      \coordinate (shift_Xin) at (0,0);
      \coordinate (shift_Hsamp) at (0,-3);
      \coordinate (shift_Hyb) at (0,-6);

      \begin{scope}[shift=(shift_Xin)]
        \draw[->] (-7.5, 0) -- (7.5, 0) node[right] {$f\text{(GHz)}$};
        \draw[->] (0, -.1) -- (0,1.6) node[above] {${X(f)}$};
        \draw (-7,.1)-- ++(0,-0.2) node[below]{-7} ++(1,0.2)-- ++(0,-0.2) node[below]{-6} ++(1,0.2)-- ++(0,-0.2) node[below]{-5} ++(1,0.2)-- ++(0,-0.2) node[below]{-4} ++(1,0.2)-- ++(0,-0.2) node[below]{-3} ++(1,0.2)-- ++(0,-0.2) node[below]{-2} ++(1,0.2)-- ++(0,-0.2) node[below]{-1} ++(1,0.2)-- ++(0,-0.2) node[below]{0} ++(1,0.2)-- ++(0,-0.2) node[below]{1} ++(1,0.2)-- ++(0,-0.2) node[below]{2} ++(1,0.2)-- ++(0,-0.2) node[below]{3} ++(1,0.2)-- ++(0,-0.2) node[below]{4} ++(1,0.2)-- ++(0,-0.2) node[below]{5} ++(1,0.2)-- ++(0,-0.2) node[below]{6} ++(1,0.2)-- ++(0,-0.2) node[below]{7};
        \draw (-.1,1)node[above left]{1}--++(.2,0);

        \draw[line width = 2pt,green](-5,0)--++(0.5,1)--++(9,0)--++(0.5,-1);

        \draw[line width = 2pt, dashed](5,1.2)node[above right]{$+f_\text{s}$} --++(0,-1.3) ;
        \draw[line width = 2pt, dashed](-5,1.2)node[above left]{$-f_\text{s}$} --++(0,-1.3) ;
      \end{scope}

      \begin{scope}[shift=(shift_Hsamp)]
        \draw[->] (-7.5, 0) -- (7.5, 0) node[right] {$f\text{(GHz)}$};
        \draw[->] (0, -.1) -- (0,1.6) node[above] {${Y_1(f)}$};
        \draw (-7,.1)-- ++(0,-0.2) node[below]{-7} ++(1,0.2)-- ++(0,-0.2) node[below]{-6} ++(1,0.2)-- ++(0,-0.2) node[below]{-5} ++(1,0.2)-- ++(0,-0.2) node[below]{-4} ++(1,0.2)-- ++(0,-0.2) node[below]{-3} ++(1,0.2)-- ++(0,-0.2) node[below]{-2} ++(1,0.2)-- ++(0,-0.2) node[below]{-1} ++(1,0.2)-- ++(0,-0.2) node[below]{0} ++(1,0.2)-- ++(0,-0.2) node[below]{1} ++(1,0.2)-- ++(0,-0.2) node[below]{2} ++(1,0.2)-- ++(0,-0.2) node[below]{3} ++(1,0.2)-- ++(0,-0.2) node[below]{4} ++(1,0.2)-- ++(0,-0.2) node[below]{5} ++(1,0.2)-- ++(0,-0.2) node[below]{6} ++(1,0.2)-- ++(0,-0.2) node[below]{7};
        \draw (-.1,1)node[above left]{1}--++(.2,0);

        \draw[line width = 2pt,green](-5,0)--++(0.5,1)--++(9,0)--++(0.5,-1);
        \draw[line width = 2pt,green](5,0)--++(0.5,1)--++(1.5,0);
        \draw[line width = 2pt,green](-5,0)--++(-0.5,1)--++(-1.5,0);

        \draw[line width = 2pt,blue, dashed](0,0)--++(-0.5,1)--++(-6.5,0);
        \draw[line width = 2pt,blue, dashed](0,0)--++(0.5,1)--++(6.5,0);

        \draw[line width = 2pt, dashed](5,1.2)node[above right]{$+f_\text{s}$} --++(0,-1.3) ;
        \draw[line width = 2pt, dashed](-5,1.2)node[above left]{$-f_\text{s}$} --++(0,-1.3) ;

        \fill[black] (7,.5) circle (0.05);
        \fill[black] (7.2,.5) circle (0.05);
        \fill[black] (7.4,.5) circle (0.05);

        \fill[black] (-7,.5) circle (0.05);
        \fill[black] (-7.2,.5) circle (0.05);
        \fill[black] (-7.4,.5) circle (0.05);
      \end{scope}

      \begin{scope}[shift=(shift_Hyb)]
        \draw[->] (-7.5, 0) -- (7.5, 0) node[right] {$f\text{(GHz)}$};
        \draw[->] (0, -.1) -- (0,1.6) node[above] {${Y_2(f)}$};
        \draw (-7,.1)-- ++(0,-0.2) node[below]{-7} ++(1,0.2)-- ++(0,-0.2) node[below]{-6} ++(1,0.2)-- ++(0,-0.2) node[below]{-5} ++(1,0.2)-- ++(0,-0.2) node[below]{-4} ++(1,0.2)-- ++(0,-0.2) node[below]{-3} ++(1,0.2)-- ++(0,-0.2) node[below]{-2} ++(1,0.2)-- ++(0,-0.2) node[below]{-1} ++(1,0.2)-- ++(0,-0.2) node[below]{0} ++(1,0.2)-- ++(0,-0.2) node[below]{1} ++(1,0.2)-- ++(0,-0.2) node[below]{2} ++(1,0.2)-- ++(0,-0.2) node[below]{3} ++(1,0.2)-- ++(0,-0.2) node[below]{4} ++(1,0.2)-- ++(0,-0.2) node[below]{5} ++(1,0.2)-- ++(0,-0.2) node[below]{6} ++(1,0.2)-- ++(0,-0.2) node[below]{7};
        \draw (-.1,1)node[above left]{1}--++(.2,0);

        \draw[line width = 2pt,green](-5,0)--++(0.5,1)--++(9,0)--++(0.5,-1);
        \draw[line width = 2pt,green](5,0)--++(0.5,1)--++(1.5,0);
        \draw[line width = 2pt,green](-5,0)--++(-0.5,1)--++(-1.5,0);

        \draw[line width = 2pt,blue, dashed](0,0)--++(-0.5,-1)--++(-6.5,0);
        \draw[line width = 2pt,blue, dashed](0,0)--++(0.5,-1)--++(6.5,0);

        \draw[line width = 2pt, dashed](5,1.2)node[above right]{$+f_\text{s}$} --++(0,-2.3) ;
        \draw[line width = 2pt, dashed](-5,1.2)node[above left]{$-f_\text{s}$} --++(0,-2.3) ;

        \fill[black] (7,.5) circle (0.05);
        \fill[black] (7.2,.5) circle (0.05);
        \fill[black] (7.4,.5) circle (0.05);

        \fill[black] (-7,.5) circle (0.05);
        \fill[black] (-7.2,.5) circle (0.05);
        \fill[black] (-7.4,.5) circle (0.05);
      \end{scope}

    \end{tikzpicture}
  }
  \caption{Interleaved sampling: Input and output spectra of two interleaved \glsfmtshort{ADC} channels~\cite[p. 5]{InterveavedADCs}.}
  \label{fig:Interleaved}
\end{figure}

Instead of partitioning the signal into multiple bands, its spectrum can also be continuously reconstructed from two converter channels.
Performing real-valued, but time shifted sampling on two or more bonded channels, the original instantaneous bandwidth can be multiplied by the number of channels by simply multiplexing between the converter channels' samples.
Reconstruction by channel multiplexing can be applied if the individual channels' time shifts result in uniform sampling over all channels.
The occurring aliasing for each individual channel will perfectly cancel out, as Fig.~\ref{fig:Interleaved} shows for a two channel setup.
Compared to the \ac{I/Q} setup, this realization requires only a single anti-aliasing filter and is therefore able to reach a slightly higher bandwidth efficiency due to less transition bands consuming bandwidth.

Considering $y_1(t)$ and $y_2(t)$, the original and the time shifted sampling, their frequency representations reveal the principle of the reconstruction:\footnote{The $*$ operator denotes a convolution. $T_\text{s}$ is the sampling interval and $f_\text{s}$ the sampling frequency.}
\begin{align}
  \label{eq:ADC1time}y_\text{1}(t)&= x(t) \sum_{k} \delta\left(t-k T_\text{s}\right)\\
  \label{eq:ADC2time}y_\text{2}(t)&= x(t) \sum_{k} \delta\left(t-k T_\text{s}- \frac{T_\text{s}}{2}\right) \\
  \label{eq:ADC1freq}Y_\text{1}(f)& = \frac{1}{T_\text{s}} X(f)  *\sum_{k} \delta\left(f-k f_\text{s}\right)\\
  \label{eq:ADC2freq}Y_\text{2}(f)&= \frac{1}{T_\text{s}} X(f) *\sum_{k} \delta\left(f-k f_\text{s}\right)\text{\,e}^{-\text{j}2\pi f T_\text{s}}
\end{align}

Due to the time shifted sampling instants, the Dirac pulses' sign will alternate for $Y_\text{2}(f)$.
Summing up both channels' frequency representations therefore leads to perfect alias cancellation in the ideal case.
Summation or multiplexing in the time domain will effectively double the sampling rate, which is equivalent.

Realizations shifting the sampling time instants or ones delaying the signal are equivalent.
With respect to a real implementation, adjustments to the shift might be necessary at runtime due to the setup's non-idealities.
Since direct \ac{RF} sampling is timing critical, the sampling clock should not be changed once calibrated.
Delaying the signal via a programmable delay network with an appropriate step size presents an applicable solution~\cite[p. 9]{InterveavedADCs}.\footnote{A fixed delay line would suffice for a specific sample rate, but is not able to compensate for subsample delay variations caused by non-idealities and temperature drift.}

\subsection{Impact of Non-idealities}
Real implementations' non-ideal hardware properties degrade the achievable reconstruction performance.
The effect is briefly derived in the following.

The approaches' respective reconstruction operations base on ideal frequency responses.
While the desired spectrum is boosted with the reconstruction gain, the image components of both channels perfectly cancel out in theory.
In the real case, the resulting desired spectrum $Y_{\text{desired}}(f)$ is superimposed by the remaining image components $Y_{\text{image}}(f)$.
Upon this, the dependence between deviations and \ac{IRR} can be derived as shown:
\begin{equation}\label{eq:intIRR}
  \begin{split}
    \text{IRR}(f)&= \frac{\lvert Y_{\text{image}}(f)\rvert}{\lvert Y_{\text{desired}}(f)\rvert} \\
    &= \frac{\lvert H_\text{c}(f) \cdot X(f) - H_\text{c}(f) \cdot H_\text{d}(f) \cdot X(f) \rvert}{\lvert H_\text{c}(f) \cdot X(f) + H_\text{c}(f) \cdot H_\text{d}(f) \cdot X(f)\rvert} \\
    &= \frac{\lvert 1 - H_\text{d}(f) \rvert}{\lvert 1 + H_\text{d}(f) \rvert}
  \end{split}
\end{equation}

Let thereby $H_1(f) = H_\text{c}(f)$ be the first channel's frequency response and the reference. 
The second channel can then be expressed as $H_2(f) = H_\text{c}(f) \cdot H_\text{d}(f)$ as common part $H_\text{c}(f)$ and $H_\text{d}(f)$ the deviation from the reference and the ideal case.
A common deviation will obviously affect the reconstructed signal, but as (\ref{eq:intIRR}) shows, the achievable image rejection only depends on the difference term $H_\text{d}(f)$ defining the relation between both channels.

This implies that the realizable \ac{IRR} is only defined by the not corrected inter-channel phase and amplitude balances, which therefore have to be minimized at any cost.
Based on (\ref{eq:intIRR}) and \ac{VNA} measurements of the analog front-ends, the approaches' individual achievable \ac{IRR} performance\footnote{For uncorrected front-end properties.} is calculated as reference for the tests presented below.

To increase the achievable \ac{IRR}, error correction can be introduced.
Frequency independent errors affect a signal's desired spectral component and the superimposed image component in the same way.
Hence, a single processing step is able to gain both ideal alias cancellation and ideal reconstruction gain.
In contrast, frequency dependent errors in general affect a signal's desired spectral component different than the superimposed image component. 
To be capable to operate on signals exceeding the Nyquist bandwidth per channel, the more complex correction scheme must be aware of the occurred aliasing.

\linespread{.972}\selectfont

\section{Practical Realization}
\subsection{Receiver Back-End}
An \ac{FPGA} design was devised enabling the Xilinx RFSoC ZU48DR to be used as receiver back-end as illustrated in Fig.~\ref{fig:targetsystemblocks}~\cite{ds889}.
Internally, realizing a modularly designed datapath, it utilizes and connects the \ac{RF} data converter to the \ac{CMAC} to realize the link to a host system.
It is natively capable of continuous sample streaming for a single converter at 5\,GSa/s.
Furthermore, the implemented buffering allows bursted sample streaming for arbitrary channel configurations.
An 100\,Gbit/s optical \ac{QSFP} is used to establish the link to the host sytem.
Details for both the protocol and the design are beyond this paper's scope.

\subsection{Analog Front-End}
Due to not length matched RF traces on the used \ac{SoC} carrier board HTG-ZRF8~\cite{HTG-Site}, suitable channel combinations are initially determined.\footnote{Length matched \ac{RF} traces would simplify processing and reconstruction of bonded channels and preserve signal quality for both in-line and post-processing.}
The \ac{I/Q} frontend is realized using a Pasternack PE2058 hybrid coupler.
Utilizing the Mini-Circuits ZFRSC-42-S+ power splitter, the interleaving approach is set up.\footnote{Filtering can be omitted for both approaches due to appropriate signal generation.}
Regarding gain and inter-channel subsample timing, a beyond Nyquist correction scheme, which is also out of scope of the present paper, is applied in post-processing.
 
\begin{figure}[h]
  \centering  
  \vspace{-.3cm}
  \resizebox*{.9\linewidth}{!}{\input{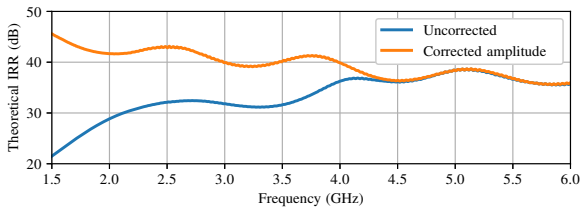}}
  \vspace{-.3cm}
  \caption{\glsfmtshort{I/Q} sampling: Theoretical \glsfmtshort{IRR} with and without correction of the amplitude imbalance.}\label{fig:VNAHybridIRR}
\end{figure}

\begin{figure}[h]
  \centering
  \vspace{-.3cm}
  \resizebox*{.9\linewidth}{!}{\input{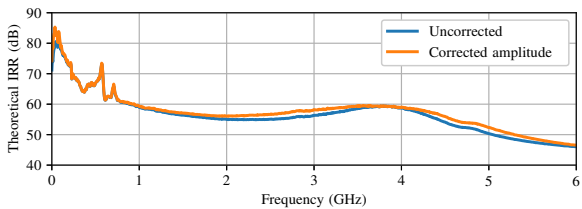}}
  \vspace{-.3cm}
  \caption{Interleaved sampling: Theoretical \glsfmtshort{IRR} with and without correction of the amplitude imbalance.}\label{fig:VNASplitIRR}
\end{figure}

Based on \ac{VNA} measurements of the utilized hybrid and splitter hardware, the \ac{IRR} achievable without calibration and correction can be determined using (\ref{eq:intIRR}).
Figures~\ref{fig:VNAHybridIRR} and \ref{fig:VNASplitIRR} illustrate the resulting limits for both front-ends for fully uncorrected and amplitude imbalance  corrected properties.

\section{Results} 
The applicability and reconstruction benefit of the introduced bandwidth extension techniques together with the Xilinx RFSoC platform shall be determined for both aliased and non-aliased cases.
Periodic and complex baseband input signals with 80 and 400\,MHz bandwidth are generated by an arbitrary waveform signal generator at several critical frequencies of both architectures.

\subsection{I/Q Sampling}
\begin{figure}[h]
  \centering
  \vspace{-.3cm}
  \resizebox*{.9\linewidth}{!}{\input{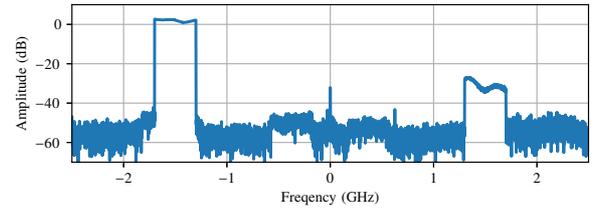}}
  \vspace{-.3cm}
  \caption{I/Q Sampling: Spectrum of the reconstructed 400\,MHz input signal at carrier 3.5\,GHz. 33\,dB \ac{IRR} is realized.}\label{fig:HybSpectrum}
\end{figure}

After the reconstruction, the \ac{IRR} can directly be determined from the non-aliased spectrum.
Figure~\ref{fig:HybSpectrum} exemplarily shows the spectrum of the analytic signal reconstructed from the measurement of an input signal at carrier 3.5\,GHz.\footnote{In contrast to Fig.~\ref{fig:hyb90rxanalytic}, the spectrum is periodified due to sampling and the bands therefore appear shifted.}
Comparing the bands at the center frequencies $\pm 1.5$\,GHz, the achieved \ac{IRR} of 33\,dB can easily be identified.

Table~\ref{tab:IRRHybrid} states the achieved \ac{IRR} values for input signals with 400\,MHz bandwidth.
Observing the impact of the applied post-processing, the significance of calibration and correction becomes evident: 
Up to 17\,dB of total improvement were achieved.

Reaching up to 36\,dB \ac{IRR}, the approaches' performance was improved even further for 80\,MHz wide input signals. 
Overall, the post-processed measurements approached the theoretical limits derived from a \ac{VNA} measurement of the PE2058 within 4 to 8\,dB.
Hereby, no calibration data was included for correction.

\begin{table}[!h]
  \centering 
  \caption{\glsfmtshort{I/Q} sampling: Realized \glsfmtshort{IRR} for input signals with 400\,MHz bandwidth at different carrier frequencies with and without post-processing.}
  \label{tab:IRRHybrid}
  \resizebox*{.85\linewidth}{!}{
  \begin{tabular}{|L{.2\linewidth}||L{.2\linewidth}|L{.2\linewidth}|}
    \hline
    \multicolumn{1}{|l||}{\multirow{2}{.19\linewidth}{\parbox{\linewidth}{\textbf{Input signal\\carrier (GHz)}}}} & \multicolumn{2}{c|}{\textbf{Realized \glsfmtshort{IRR} (dB)}}                                                                     \\\cline{2-3}
                                                                 & \textbf{Raw}                                                & \textbf{Post-processed } \\
    \hline
    \hline
    2.25                                                         & 18.252                                                       & 30.359         \\
    \hline
    2.325                                                        & 17.883                                                       & 32.482         \\
    \hline
    3.5                                                          & 16.665                                                       & 33.904         \\
    \hline
    5.125                                                        & 17.697                                                      & 21.873         \\
    \hline
    5.75                                                         & 13.411                                                       & 30.855         \\
    \hline
  \end{tabular}}
  
  \vspace{-.2cm}
\end{table}

To evaluate the reconstruction's benefit in case of in-band interference, the \ac{SINR} was analyzed. 
Therefore, carriers were chosen at integer multiples of the Nyquist frequency such that the signal fully aliases with itself.
The \ac{SINR} improvement due to the reconstruction was then determined from the impulse responses derived by deconvolution.
Figure~\ref{fig:HybridAliasedReconstruction} shows the impulse responses before and after reconstruction.
As expected, the reconstruction gain of 6\,dB was observed as well as an \ac{SINR} improved by approximately 20\,dB.

\begin{figure}
  \centering
  \vspace{-.1cm}
  \begin{subfigure}[b]{.45\linewidth}
    \centering
    \resizebox*{\linewidth}{!}{\input{PIC/2G5HybridAliasedImpulseResponse.pgf}}
    \vspace{-.7cm}
    \caption{Individual channels: Input signal carrier at 2.5\,GHz.}\label{fig:2G5HybridChannels}
  \end{subfigure}
  \begin{subfigure}[b]{.45\linewidth}
    \centering
    \resizebox*{\linewidth}{!}{\input{PIC/2G5HybridAliasedImpulseResponseCombined.pgf}}
    \vspace{-.7cm}
    \caption{Combined channels: Input signal carrier at 2.5\,GHz.}\label{fig:2G5HybridReconstructed}
  \end{subfigure}
  \vspace{-.1cm}
  \caption{\glsfmtshort{I/Q} sampling: Reconstruction of channel impulse response using fully aliased signals with 80\,MHz bandwidth.}		\label{fig:HybridAliasedReconstruction}
  \vspace{-.3cm}
\end{figure}

\subsection{Interleaved Sampling}
Evaluating the reconstruction for the interleaving front-end architecture, \ac{IRR} values of up to 49\,dB were achieved for 80\,MHz wide input signals.
As Table~\ref{tab:IRRInterleaving} shows, \ac{IRR} values beyond 10\,dB were realized for 400\,MHz wide signals in the first Nyquist zone.

In contrast, a severe degradation occured for signals in the second Nyquist zone and no significant \ac{IRR} values were achieved.
Comparing the individual channels' spectra in amplitude and phase, strongly increasing channel imbalance was found for high frequencies.
This effect might be caused by trace and connector realizations differing between channels on the \ac{SoC} carrier board and requires further investigation. 

\begin{table}[h]
  \centering
  \caption{Interleaved sampling: Realized \glsfmtshort{IRR} for input signals with 400\,MHz bandwidth at different carrier frequencies with and without post-processing.}
  \label{tab:IRRInterleaving}
  \resizebox*{.85\linewidth}{!}{
    \begin{tabular}{|L{.2\linewidth}||L{.2\linewidth}|L{.2\linewidth}|}
      \hline
      \multicolumn{1}{|l||}{\multirow{2}{.19\linewidth}{\parbox{\linewidth}{\textbf{Input signal\\carrier (GHz)}}}} & \multicolumn{2}{c|}{\textbf{Realized \glsfmtshort{IRR} (dB)}}                                                                                                       \\\cline{2-3}
                                                                   & \textbf{Raw}                                                &  \textbf{Post-processed} \\
      \hline
      \hline
      1.0                                                          & 12.330                                                       & 23.014          \\
      \hline
      2.325                                                        & 3.877                                                        & 12.797         \\
      \hline
      3.0                                                          & -0.047                                                       & 8.370         \\
      \hline
      4.0                                                          & -4.685                                                       & 2.686          \\
      \hline
    \end{tabular}}
    \vspace{-.2cm}
\end{table}

Examining the reconstruction's effect for a fully aliased signal centered at the Nyquist frequency -- similar to the hybrid approach -- a 6\,dB reconstruction gain was observed as well as an \ac{SINR} improved by 15\,dB.

Combining the broadband timing correction scheme with an iterative optimizer, minimizing the autocorrelation-based subsample timing error measure intruduced by~\cite{InterveavedADCs}, allowed fully correcting the sample timing for this approach without any a priori information.
Additionally, utilizing effective power equalization, no a priori information was required for post-processing at all.

\section{Conclusions}
The Xilinx RFSoC platform was successfully utilized to realize a modular, versatile, and scalable receiver solution.
It implements a direct \ac{RF} receiver with channel bonding for bandwidth extension beyond Nyquist and reliable sample streaming for multiple coherent 5\,GSa/s channels over 100\,Gbit/s Ethernet.
Both introduced and realized front-end architectures were demonstrated to be capable of reassembling 5\,GHz of continuous bandwidth from two converter channels at 5\,GSa/s.
While in theory, both realize a flat frequency response without incurring aliasing, the real implementations suffered from non-idealities.
By reaching 36\,dB \ac{IRR}, the realized \ac{I/Q} sampling approach was able to approach its theoretical limit to up to 4\,dB.
The interleaved sampling approach even realized \ac{IRR} values in excess of 40\,dB, but was strongly affected by degradation for high frequencies and stays well below its theoretical limit.
The utilized broadband error correction schemes were able to significantly improve both front-ends' performance, without including channel calibration data.
For the interleaving setup, they even allowed to optimize the correction parameters without any a priori information.
A detailed, theoretical derivation of the correction algorithm was out of the scope and will be given in a future paper.

To further enhance the achitectures' capabilities and resolve the occurred degradation issues, the properties of the converters, the \ac{PCB}, and the front-ends' components have to be investigated and included for correction.
An ideal calibration measurement then covers the whole channel with its complex frequency response, while any parameter drift is to be minimized after calibration.

\section*{Acknowledgment}
The authors acknowledge the financial support by the Federal Ministry of Education and Research of Germany in the project "ICAS4Mobility" (grant number: 16KISK241) and by the federal state of Thuringia in the project "KREAT\"OR" (grant number: 2021 FGI 007).

\linespread{.96}\selectfont

\bibliographystyle{IEEEtran}
\bibliography{paper.bib}

\end{document}